\documentstyle[]{article}
\input psfig
\def\hybrid{\topmargin 0pt      \oddsidemargin 0pt
        \headheight 0pt \headsep 0pt
        \textwidth 6.5in        
        \textheight 9.0in         
        \marginparwidth 0.0in
        \parskip 5pt plus 1pt   \jot = 1.5ex}
\catcode`\@=11
\def\marginnote#1{}

\newcount\hour
\newcount\minute
\newtoks\amorpm
\hour=\time\divide\hour by60
\minute=\time{\multiply\hour by60 \global\advance\minute by-\hour}
\edef\standardtime{{\ifnum\hour<12 \global\amorpm={am}%
        \else\global\amorpm={pm}\advance\hour by-12 \fi
        \ifnum\hour=0 \hour=12 \fi
        \number\hour:\ifnum\minute<10 0\fi\number\minute\the\amorpm}}
\edef\militarytime{\number\hour:\ifnum\minute<10 0\fi\number\minute}

\def\draftlabel#1{{\@bsphack\if@filesw {\let\thepage\relax
   \xdef\@gtempa{\write\@auxout{\string
      \newlabel{#1}{{\@currentlabel}{\thepage}}}}}\@gtempa
   \if@nobreak \ifvmode\nobreak\fi\fi\fi\@esphack}
        \gdef\@eqnlabel{#1}}
\def\@eqnlabel{}
\def\@vacuum{}
\def\draftmarginnote#1{\marginpar{\raggedright\scriptsize\tt#1}}

\def\draft{\oddsidemargin -.5truein
        \def\@oddfoot{\sl preliminary draft \hfil
        \rm\thepage\hfil\sl\today\quad\militarytime}
        \let\@evenfoot\@oddfoot \overfullrule 3pt
        \let\label=\draftlabel
        \let\marginnote=\draftmarginnote
   \def\@eqnnum{(\theequation)\rlap{\kern\marginparsep\tt\@eqnlabel}%
\global\let\@eqnlabel\@vacuum}  }

\def\numberbysection{\@addtoreset{equation}{section}
        \def\theequation{\thesection.\arabic{equation}}}
\numberbysection

\renewcommand{\theequation}{\thesection.\arabic{equation}}
\def\titlepage{\@restonecolfalse\if@twocolumn\@restonecoltrue\onecolumn
     \else \newpage \fi \thispagestyle{empty}\c@page\z@
        \def\thefootnote{\fnsymbol{footnote}} }

\def\endtitlepage{\if@restonecol\twocolumn \else  \fi
        \def\thefootnote{\arabic{footnote}}
        \setcounter{footnote}{0}}  
\relax

\hybrid

\makeatletter
\newdimen\normalarrayskip              
\newdimen\minarrayskip                 
\normalarrayskip\baselineskip
\minarrayskip\jot
\newif\ifold             \oldtrue            \def\new{\oldfalse}
\def\arraymode{\ifold\relax\else\displaystyle\fi} 
\def\eqnumphantom{\phantom{(\theequation)}}     
\def\@arrayskip{\ifold\baselineskip\z@\lineskip\z@
     \else
     \baselineskip\minarrayskip\lineskip2\minarrayskip\fi}

\def\@arrayclassz{\ifcase \@lastchclass \@acolampacol \or
\@ampacol \or \or \or \@addamp \or
   \@acolampacol \or \@firstampfalse \@acol \fi
\edef\@preamble{\@preamble
  \ifcase \@chnum
     \hfil$\relax\arraymode\@sharp$\hfil
     \or $\relax\arraymode\@sharp$\hfil
     \or \hfil$\relax\arraymode\@sharp$\fi}}

\def\@array[#1]#2{\setbox\@arstrutbox=\hbox{\vrule
     height\arraystretch \ht\strutbox
     depth\arraystretch \dp\strutbox
     width\z@}\@mkpream{#2}\edef\@preamble{\halign \noexpand\@halignto
\bgroup \tabskip\z@ \@arstrut \@preamble \tabskip\z@ \cr}%
\let\@startpbox\@@startpbox \let\@endpbox\@@endpbox
  \if #1t\vtop \else \if#1b\vbox \else \vcenter \fi\fi
  \bgroup \let\par\relax
  \let\@sharp##\let\protect\relax
  \@arrayskip\@preamble}
\def\eqnarray{\stepcounter{equation}%
              \let\@currentlabel=\theequation
              \global\@eqnswtrue
              \global\@eqcnt\z@
              \tabskip\@centering
              \let\\=\@eqncr
              $$%
 \halign to \displaywidth\bgroup
    \eqnumphantom\@eqnsel\hskip\@centering
    $\displaystyle \tabskip\z@ {##}$%
    &\global\@eqcnt\@ne \hskip 2\arraycolsep
         $\displaystyle\arraymode{##}$\hfil
    &\global\@eqcnt\tw@ \hskip 2\arraycolsep
         $\displaystyle\tabskip\z@{##}$\hfil
         \tabskip\@centering
    &{##}\tabskip\z@\cr}
\makeatother

\def\bea{\begin{eqnarray}}
\def\eea{\end{eqnarray}}
\def\nn{\nonumber}

\def\beq{\begin{equation}}
\def\eeq{\end{equation}}
\def\be{\beq\new\begin{array}{c}}
\def\ee{\end{array}\eeq}

\def\Tr{{\rm Tr}}

\def\Im{{\rm Im}}

\def\2{{1\over 2}}

\begin{document}

\begin{titlepage}
\setcounter{footnote}0
\begin{center}
\hfill FIAN/TD-7/97\\
\hfill ITEP/TH-24/97\\
\hfill hep-th/9706050\\

\vspace{0.3in}
{\LARGE\bf  Insights and Puzzles from Branes: $4d$ SUSY
Yang-Mills from $6d$ Models}
\\
\bigskip\bigskip\bigskip

{\Large A.Marshakov}
\footnote{E-mail address:
mars@lpi.ac.ru, andrei@heron.itep.ru, marshakov@nbivms.nbi.dk},
\\
{\it Theory Department, P. N. Lebedev Physics
Institute, Leninsky prospect 53, Moscow, ~117924, Russia\\
and\\ ITEP, Moscow, ~117259, Russia}\\
\bigskip
{\Large M.Martellini}
\footnote{Landau Network at Centro Volta, Como, Italy}\\
{\it Dipt. di Fisica, Univ. of Milano and I.N.F.N. Sez. di
Milano, Via Celoria 16, 20133 Italy}\\
and\\
{\Large A.Morozov}
\footnote{E-mail address:
morozov@vxdesy.desy.de}
\\
{\it ITEP, Moscow, ~117 259, Russia}\\
\end{center}
\bigskip \bigskip

\begin{abstract}
$5$-branes of nontrivial topology are associated in the
Diaconescu-\-Hanany-\-Witten-\-Witten (DHWW) approach with
the Seiberg-\-Witten (SW) theory of low-energy
effective actions. There are two different "pictures",
related to the IIA and IIB phases of $M$-theory.
They differ by the choice of $6d$ theory on the $5$-brane
world volume. In the IIB picture it is just the $6d$ SUSY
Yang-Mills, while in the IIA picture it is a theory of SUSY
self-dual 2-form. These two pictures appear
capable to describe the (non-abelian) Lax operator and
(abelian) low-energy effective action respectively.
Thus IIB-IIA duality is related to the duality between
Hitchin and Whitham integrable structures.
\end{abstract}

\end{titlepage}

\newpage
\setcounter{footnote}0

\section{Introduction}

    According to general principles of string program,
various quantum field theory models are identified with the
various classical configurations (``vacua'') of the
string theory, which can be considered as a kind of universal
object of the Quantum Field Theory.

    Recent advances in this direction appeared due to the
introduction of the new class of string vacua:
described in terms of the BPS-saturated branes.
At present stage the focus of research is on the branes
of non-trivial topology. In particular, the system of
parallel $p$-branes appears to be associated with the
$p+1$-dimensional (SUSY) Yang-Mills models \cite{W1}
and non-perturbative phenomena in the Yang-Mills theory can be
reformulated as interactions of branes.
This interaction makes the
geometry of branes non-flat and in the low-energy limit the
nontrivial geometry plays the same role as compactification, thus
effectively reducing the naive number of the space-time dimensions.
It opens a way for geometrical reinterpretation of the interaction
in Yang-Mills theory: the old dream is getting real.

The simplest realization of this idea -- the DHWW
construction \cite{D,HW,W2} allows one to
associate the non-perturbative low-energy $p$-dimensional
SUSY Yang-Mills (SYM) theory with non-trivial vacua of the
$p+1$-dimensional SUSY gauge theory of forms on the
brane world volume
\footnote{The role of non-trivial SYM vacua was
emphasised in \cite{W1}. Technically we seem to overcome
the argument of \cite{W1} against considering the Dirichlet
$5$-branes (i.e. against our type-IIB picture) due to
nontrivial boundary conditions for the scalar fields, see
sect.4 below.}.
This sheds light to the mysteries of the SW theory
\cite{SW1,SW2,SW3} of the low-energy effective
actions (RG-flows) in $N=2$ SYM theory.
In \cite{W2} Witten interpreted the SW curve
as topologically nontrivial constituent of the brane
configuration. In this paper we discuss how the
(Toda chain) Lax operator may arise from the DHWW construction
for $p=4$. As anticipated in \cite{MW2,G}
this can be understood in
terms of $p+1=6$-dimensional SUSY Yang-Mills theory
on the world volume of the $5$-brane.
While being adequate for the description of renormalization
group (RG) flow from a non-abelian theory in the ultraviolet (UV)
limit and of emerging integrable structure,
such essentially IIB stringy picture is not enough, however, to
obtain a simple description of the prepotential.
Instead, this is straightforward in the dual IIA-inspired
picture \cite{W2} when the SUSY $p+1=6$ dimensional
theory on a brane world volume is a theory of self-dual
2-form.

The whole construction is a direct
developement of Diaconescu's \cite{D} original reasoning
for $p=1$ (and its analog for $p=3$), where the Nahm equations
\cite{N,CG} arise as non-trivial generalization of the
Toda-chain formalism (relevant for $p=4$).
It makes the appearence of a complex spectral curve
and prepotential a little more natural --
though it is hardly an explanation in intrinsic terms
of the SYM theory, and universality of emerging structures
is not quite obvious.
Even more important, these two ingredients of the SW
theory (the curve and the prepotential for the {\it given}
curve) remain linked to two different "pictures"
in M-theory.
Thus the main dynamical question --
of the derivation of the SW ansatz {\it as a whole} --
or \cite{GKMMM,oth}, of the derivation of
abelian Whitham effective low-energy dynamics
from the non-abelian Yang-Mills-Hitchin one --
is not resolved, but reinterpreted as the
question about duality between IIA and IIB-type pictures,
i.e. is put closer to the main stream of the studies
of string dualities.

\section{DHWW construction}

In the DHWW construction one essentially considers a $5$-brane in
$M$-theory
\footnote{Original papers \cite{D} and \cite{HW} deal with
various degenerations of this construction,
when some compactification radia go to zero,
thus giving rise to $1,3,4$-branes.
See also \cite{pre} for some important preliminary works and
\cite{dev} for more examples.}.
In the first-quantized formalism (still the only
one available in most string theory considerations),
its dynamics is effectively described by the world-volume
($6d$) theory of either the SYM -- in the type
IIB picture or the SUSY self-dual 2-form
$C = \{C_{MN}\}$, $dC = \ast dC$ -- in the type IIA picture.
After compactification on a circle the low-energy theory
is $5d$ SYM, and
compactification on a Riemann surface (complex curve)
$\Sigma$ leads to a world-volume $4d$ $N=2$ SYM.

As usual \cite{W1}, the gauge group $SU(N_c)$ is defined by
topology of the brane, and in the low-energy regime it is
broken down to the abelian $U(1)^{N_c-1}$,
with the scalar (adjoint Higgs) vacuum expectation values
identified with (some) moduli of the
complex structure on $\Sigma$.
Most important, in order to allow interpretation in terms of
spontaneously broken $SU(N_c)$ gauge symmetry, the choice
of the Riemann surface $\Sigma$ is severely
restricted: to hyperelliptic complex curves,
being at the same time
$N_c$-fold coverings of a cylinder
and associated to the Toda-chain integrable systems.
Their appearance in the form of either $2$-fold or
$N_c$-fold covering is responsible for two possible
descriptions: in terms of $SU(2)$ and $SU(N_c)$ groups --
well known both in the brane language
\cite{HW,W2,dev} and in the
approach based on integrable systems \cite{GMMM}.

The DHWW construction is essentially as follows \cite{W2}:
one starts with embedding the $5$-brane's world-volume
into $11$-dimensional target space-time with the co-ordinates
$x^0,...,x^{10}$. We further assume that the target space has
topology $R^{9}\times T^{2} = R^{9}\times S^1\times S^1$.
The second $S^1$ (spanned by $x^{10}$)
will be ignored in what follows\footnote{
Of course, it is crucially important for accurate
embedding of IIA and IIB strings into
generic $M$-theory frame and for the explanation
of the {\it origins} of the two pictures
and their interrelation -- this is however beyond
the scope of present paper.},
while the essential compact co-ordinate will be called $x^9$.
Now, one proceeds with a $5$-brane with the world-volume
topology $R^5\times S^1$ and
parameterized by $(x^0,x^1,x^2,x^3,x^6,x^{9})$,
leaves aside four flat dimensions ($x^0,x^1,x^2,x^3$ --
the space-time of the low-energy  $4d$ $N=2$ SYM
theory), and ends up with a cylinder $R\times S^1$ embedded
into the target space along $(x^6,x^9)$ dimensions.
We use the notation $z = x^6+ix^9$ for the
corresponding complex co-ordinate.
Next, in order to get a non-trivial gauge group:
spontaneously broken $SU(N_c) \rightarrow U(1)^{N_c-1}$,
one needs $N_c$ {\it parallel} copies of the cylinder
(see Fig.1).
Different cylinders have different positions in
``transverse'' space $V^{\bot} = (x^4,x^5,x^7,x^8)$.
Moreover \cite{D}, to get
not just a remnant $U(1)^{N_c-1}$, but indeed a
spontaneously broken non-abelian theory, these parallel cylinders
should come from a {\it bound state} of $N_c$ branes i.e.
should be different parts of the {\it same}
brane \cite{W2}. It means that the cylinders should be
all glued together (see Fig.2).
Actually in the weak coupling limit they are glued
at infinity, while increasing coupling constant
distorts them along their entire length.
Since the cylinders are 2-dimensional and parallel, projection of
this entire configuration onto $V^{\bot}$ is also 2-dimensional.
Supersymmetry requires it to be just a plane in $V^{\bot}$:
$C = R^2 \in V^{\bot}$, we will describe it in terms of the
complex coordinate $\lambda = x^4 +ix^5$.

Introducing coordinate $w = e^{z}$ to describe a cylinder,
we see that the system of non-interacting branes (Fig.1) is
given by $z$-independent equation
\be
P_{N_c}(\lambda) = \prod_{\alpha = 1}^{N_c}
(\lambda - \lambda_\alpha) = 0,
\label{nibr}
\ee
while their bound state (Fig.2) is described by the complex curve
$\Sigma _{N_c}$ \cite{W2}:
\be
\Lambda^{N_c}\left(w + \frac{1}{w}\right) = 2P_{N_c}(\lambda)
\ \ \ \ \ \ \ \ \hbox{or}
\nn \\
\Lambda^{N_c}\cosh z = P_{N_c}(\lambda)
\label{todacur}
\ee
In the weak-coupling limit  $\Lambda \rightarrow 0$
(i.e. $\frac{1}{g^2} \sim \log\Lambda \rightarrow \infty$)
one comes back to  disjoint branes (\ref{nibr})
\footnote{In other words, for $|z| \ll |\log\Lambda|$,
$\lambda$ is almost independent of $z$ and confined
to be almost equal to
some of $\lambda_\alpha$. Only when $|z| \sim |\log\Lambda|$
coordinate $\lambda$ is allowed to deviate from fixed position and
"interpolate" between --
different $\lambda_\alpha$'s.
}
\footnote{Eq.(\ref{todacur})
and Fig.2 decribe a hyperelliptic curve --
a double covering of a punctured Riemann sphere,
\be
y^2 = \frac{\Lambda^{2N_c}}{4}\left(w - \frac{1}{w}\right)^2 =
P_{N_c}^2(\lambda) - \Lambda^{2N_c}
\nn
\ee
Such hyperelliptic curves and their period matrices
are the main ingredients of the SW ansatz \cite{KLTY,AF,HO}
for the $4d$ $N=2$ SUSY low-energy effective actions.}.

Thus we finally got a $5$-brane of topology
$R^3\times\Sigma _{N_c}$ embedded into a subspace $R^5\times S^1$
(spanned by $x^1,...,x^6,x^9$) of the full target space. The
periodic coordinate is
\be\label{perco}
x^9 = \arg P_{N_c}(\lambda) = \Im\log P_{N_c}(\lambda) =
\sum _{\alpha = 1}^{N_c} \arg (\lambda - \lambda _{\alpha})
\ee

\section{Lax operator}

According to \cite{GKMMM}, occurence of the complex curves
like $\Sigma _{N_c}$ is
a manifestation of hidden integrable structure behind
the theory of renormalization group (RG) flows.
Namely \cite{GKMMM,MW1,NT,IM3,Mar,Do,Kl},
equation (\ref{todacur}) describes
the spectral curves of the ($0+1$-dimensional, $N_c$-periodic)
Toda-chain hierarchy:
\be
\det_{N_c\times N_c}
(\Lambda{\cal L}(z) - \lambda\cdot {\bf 1}) = 0,
\label{laxtoda}
\ee
The $SU(N_c)$ Lax operator
\be\label{lax}
{\cal L}(z) = \vec p \vec H  +
e^{\vec\alpha_0\vec q}(e^{z} E_{\vec\alpha_0} +
e^{-z}E_{-\vec\alpha_0})
+ \sum_{\rm simple\ \vec\alpha >0} e^{\vec\alpha\vec q}
(E_{\vec\alpha } + E_{-\vec\alpha})
\nn \\
\vec\alpha_0 = -\sum_{simple\ \vec\alpha > 0}\vec\alpha
\ee
where $\vec H$ are the diagonal (Cartan) $SU(N_c)$ matrices
and $E_{\vec\alpha }$ are matrices corresponding to the roots of
$SU(N_c)$: $E_{\vec\alpha _{ij},mn} = \delta _{mi}\delta _{nj}$.
Only the simple roots with $j=i\pm 1$ appear in (\ref{lax}).
The Hamiltonians
of the Toda chain are symmetric polynomials of parameters
$\lambda _{\alpha}$ in (\ref{nibr}), e.g.
\be
\Lambda ^2h_2 \equiv \Lambda ^2\left( {\vec p}^2 +
e^{2\vec\alpha_0\vec q} + \sum_{\rm simple\ \vec\alpha >0}
e^{2\vec\alpha\vec q} \right)
= \sum _{\alpha < \beta}\lambda _{\alpha}\lambda _{\beta}
\ee
From the point of view of the DHWW construction the
shape of the complex curve $\Sigma_{N_c}$ should not be just
guessed or postulated:
it describes the eigenvalues of the scalar field $\Phi(z)$ --
the member of the $6d$ supermultiplet,
i.e. $\lambda _{\alpha}$'s are
solutions of the equation
\be\label{curgen}
\det_{N_c\times N_c} (\Phi(z) - \lambda\cdot {\bf 1}) = 0,
\ee
which describes mutual positions of the branes, i.e. our
cylinders. Thus, comparing (\ref{laxtoda}) and (\ref{curgen}),
we conclude that there is a natural identification
\be
\Lambda{\cal L}(z) \sim \Phi(z)
\ee
This is in fact a general point in the
Hitchin approach to integrable systems \cite{Hi,GN1,DW}
and this was already used
many times in applications of this formalism to
investigation of the Seiberg-Witten effective theory
\cite{DW,Mart,GM,IM1,G}.

Thus, now we have something to check:
the Lax operator (\ref{lax}) should naturally arise
from the equations of motion for the scalar field $\Phi(z)$.
Moreover, in order to preserve supersymmetry, it should
satisfy an even more restrictive condition:
the linear BPS-like equation.

\section{The IIB type picture}

In order to explain how it happens, let us analyze the DHWW
construction in the type IIB picture, when the theory on the
$5$-brane world volume is $6d$ SYM. The mutual position of the
cylinders on Fig.1 is described by the coordinates in orthogonal
space $V^{\bot}$ (spanned by $x^4,x^5,x^7,x^8$),
i.e. by four scalar fields
$\Phi^{(4)},\Phi^{(5)},\Phi^{(7)},\Phi^{(8)}$ -- the members
of the $6d$ SYM gauge multiplet. As usual, they are taking
values in the adjoint representation of the gauge group $SU(N_c)$,
where $N_c$ is the number of cylinders, i.e. co-ordinates
$x^4,\ldots,x^8$ are substituted by non-obligatory commuting
matrices $\Phi^{(4)},\ldots, \Phi^{(8)}$. The members of gauge
multiplet are associated with the open strings streched
between the cylinders, the corresponding $10d$ vector field
$A_S=\{ A_M,A_{\mu} \} $ in the bulk naturally decomposes
into the components with
$M=0,1,2,3,6,9$ -- considered as $6d$ vector from the point of
view of the effective theory on the brane -- and with
$\mu =4,5,7,8$
-- associated with four above-mentioned scalars.
The nonabelian interaction arises due
to the processes like in Fig. 3.

In a vacuum state the scalar fields satisfy
the BPS-like condition
\be\label{bps}
D_M\Phi \equiv \partial _M\Phi + [A_M,\Phi ] = 0,
\ \ \ F_{MN} = 0
\ee
This equation is so simple at least when only one of the fields
$\Phi^{(4)},\ldots,\Phi^{(8)}$ is nonvanishing.
This is essentially the case for the configuration of Fig.2,
arising from Fig.1 when the brane
interaction is switched on: Fig.2 implies that some scalar field,
say $\Phi\equiv \Phi^{(4)}+i\Phi^{(5)}$, develops a nonvanishing
$z$-dependent vacuum expectation value -- this is exactly the
statement that the cylinders are distorted and glued together.
In order to explain/derive Fig.2, it is necessary to demonstrate
that eq. (\ref{bps}) has a {\it non-trivial} solution $\Phi (z)\neq
const$. The reason for this is that non trivial boundary
conditions are imposed on $\Phi$ at $z\rightarrow\pm\infty$.

In order to understand how they should be adequately described,
let us consider first the UV-finite version of the SW theory and
then take the double-scaling limit back to the asymptotically free
situation.

The way to do this is well known and examined in detail
in \cite{SW2,DW,IM1}. One should add to the $4d$ $N=2$ SUSY
pure gauge theory an extra matter hypermultiplet in the adjoint
representation with the mass $m$. When $m=0$, one gets a theory
with $N=4$ SUSY which is UV-finite with the UV coupling
constant $\tau = {i\over g_{UV}^2} + {\theta\over 2\pi}$. When
$m\neq 0$ it remains UV-finite, but acquires a nontrivial
RG-flow. The original pure $N=2$ SYM theory is restored in the
double-scaling limit when $\tau\rightarrow i\infty$,
$m\rightarrow\infty$, so that $m^{N_c}e^{2\pi i\tau}\equiv\Lambda
^{N_c}$ remains finite. Within the framework of the SW theory
this corresponds to a spectral curve -- a cover of a torus
(with complex modulus $\tau $) which in the limit $\tau\rightarrow
i\infty$ degenerates into a cylinder. Associated integrable system
is the elliptic Calogero-Moser model with the coupling constant
$m$ \cite{GN2,Mart,GM,IM1} which in the double-scaling limit
\cite{Ino} turns into a Toda chain.

From the point of view of the brane picture at Fig.1 it means that
one should first substitute the cylinders by tori with the same
modulus $\tau$. The isolated tori would correspond to the
vanishing parameter $m$, while non-vanishing $m$ means that the
scalar field $\Phi$ acquires nontrivial boundary consitions, or
is a section of a nontrivial (holomorphic) bundle. In other words,
when one takes cylinders from Fig.1 and glues the ends to make a
torus -- the fields jump, and on the torus the equation
(\ref{bps}) acquires a non-zero r.h.s.,
which survives in the double-scaling limit.

More technically,
on a torus one cannot fix the gauge ${\bar A} \equiv A_6+iA_9 =0$,
by gauge transformation ${\bar A}$ can be at best
brought to diagonal form ${\bar A} = diag(a_1,...,a_{N_c})$.
Then the corresponding component of equation (\ref{bps})
becomes\footnote
{There are different ways to interpret the $\delta $-function
in the r.h.s. of (\ref{hi}): one can say, for example \cite{G},
that $z=z_0$ is the point where "vertical brane" of
original presentation of \cite{D,HW} intersect the
"horisontal branes" -- our tori.
In Fig.2 the point $z_0$ is at infinity,
i.e. exactly where nontrivial
boundary conditions are imposed in eq.(\ref{bps}).}
\be\label{hi}
\bar\partial\Phi ^{ij} + (a_i-a_j)\Phi ^{ij} =
m(1-\delta ^{ij})\delta (z - z_0)
\ee
so that
\be
\Phi ^{ij}(z) = p_i\delta ^{ij}+ m(1-\delta ^{ij})
e^{(a_i-a_j)(z-{\bar z})}
{\theta (z - z_0 + {a_i-a_j\over\pi\Im\tau})\over\theta (z - z_0)}
\ee
To compare with the conventional Lax operator of the elliptic
Calogero-Moser model \cite{KriCal},
one should make a gauge transformation
\be\label{laxkri}
\Phi ^{ij}(z)\rightarrow (U^{-1}\Phi U)^{ij}(z) =
p_i\delta ^{ij}+ m(1-\delta ^{ij})
e^{(a_i-a_j)z}
{\theta (z - z_0 + {a_i-a_j\over\pi\Im\tau})\over\theta (z - z_0)}
\ee
with $U^{ij} = e^{(a_i-a_j){\bar z}}$, then the explicit dependence
on ${\bar z}$ is eliminated but $\Phi (z)$ becomes a
multivalued function or a section of a
nontrivial bundle over torus. The Lax operator (\ref{lax})
is obtained from (\ref{laxkri}) in the double-scaling limit
\cite{Ino}.

Of course, there is a way to describe relevant nontrivial
boundary conditions directly in
terms of Fig.2 (without additional
compactification-/-decompactification of the
$x^6$ dimension), but the above
presentation reveals better the origins of what happens.

Thus, in the type IIB picture we derived the {\it shape}
of the curve (\ref{todacur}), (\ref{laxtoda}), (\ref{curgen})
"from the first principles".
The next step would be to derive the effective action of
emerging low-energy $4d$ theory.
However, here one runs into problems.

Since the world-volume action is not quadratic,
it is necessary to take non-trivial average
over the fields which become massive due to the Higgs
mechanism, moreover this average includes
non-perturbative corrections.
This is more or less the same as the original problem in the
SW theory, without any obvious simplifications.
As explained in \cite{GKMMM,Mar},
from the point of view of integrable hierarchies the
derivation of the low-energy effective action is the aim
of the so-called Bogolyubov-Whitham averaging method,
which is still far from being throughly developed.
Remarkably, despite such problems,
the net {\it result} of this procedure can be easily
described in terms of period integrals on spectral surfaces,
i.e. in the framework of the prepotential theory
(or that of the quasiclassical $\tau$-functions)
\cite{Kr,Dubr,IM2},
which is in a sense "dual" to the Hitchin theory \cite{Mar,Mar2}.

This is exactly in parallel to what happens
in SW theory: while there is no clear way to
{\it derive} effective action directly, the ansatz
can be easily suggested for what it actually is.
In other words,
despite the brane vacuum configuration is derived exactly
in the type IIB picture,
this picture is not sufficient itself for the derivation of
the effective action (at least it is not straightforward).
However, according to \cite{W2}, this problem can be
solved in the "dual" type IIA picture.

\section{IIA type picture}

In this picture instead of the $6d$ SYM
one considers a $6d$ SUSY theory of
self-dual 2-form $C = \{ C_{MN}\}$, $dC ={\ast} dC$
on the world volume of
a $5$-brane. It means, first, that instead of
attaching open strings to the $5$-brane, as in Fig.3, one has
to consider now "open" membranes, see Fig.4.

The important difference with the type IIB picture of the previous
section is that in the relevant approximation
the theory of 2-forms is
essentially abelian. Even if there are matrices $C_{MN}^{ij}$
in the adjoint representation of $SU(N_c)$ associated
with the vertical cylinders (membranes) attached between
$i$-th and $j$-th horisontal
cylinders, no nonabelian interacting theory can arise since such
interaction is inconsistent with the gauge invariance.
Only non-linear interaction of the non-minimal type can appear
-- like $\Tr(dC)^4$,
expressed through the tension of $C$.
Such terms, however, contain
higher derivatives (powers of momentum)
and they seem irrelevant in the low-energy effective actions.

This "abelian" nature of the 2-form theory makes the
description of the Lax operator
(vacuum expectation value of the scalar members
of the supermultiplet which describe the transverse
fluctuations of the $5$-brane), and thus the derivation
of the shape of the curve
$\Sigma _{N_c}$ in the type IIA picture, much less straightforward.
Instead, exactly due to the fact that the action on
(flat) world-volume is essentially quadratic
\be\label{act}
\int _{d^6x}|dC|^2 + \hbox{SUSY terms},
\ee
in this picture there are no corrections to the form of the
effective $4d$ action -- once $\Sigma _{N_c}$ is given.
It is enough to
consider the dimensional reduction of (\ref{act}) from 6 to 4
dimensions \cite{W2}.

Such reduction implies that the 2-form $C$ is decomposed as
\be\label{ver}
C_{\mu z} = \sum _{i=1}^{N_c-1}\left(A_{\mu}^i(x)d\omega _i(z) +
{\bar A}_{\mu}^i(x)d{\bar \omega} _i(\bar z)\right)
\ee
where $d\omega _i$ are canonical holomorphic 1-differentials on
$\Sigma _{N_c}$
\footnote{Actually, before the double scaling limit described
in the previous section, the curve $\Sigma _{N_c}$ is compact
of genus $N_c$ (when the $x^6$ direction is also compactified
along with $x^9$) and the curve possesses $N_c$
holomorphic differentials.
However, one of them develops a simple pole
when $z\rightarrow\infty$
and thus is ignored in (\ref{act}).
Also in our simplified description
we ignore other components of the 2-form:
$C_{z\bar z}$ and $C_{\mu\nu}$
which are related to each other by the selfduality condition
$$
\partial _{\lambda}C_{z\bar z} =
{1\over\sqrt{g}}\epsilon_{\lambda\mu\nu\rho}
\partial _{\rho}C_{\mu\nu}
$$
and correspond from the $4d$ point of view to a (real) scalar.
The whole picture thus would contain {\it three complex}
scalar fields, two of which become massive
in the configuration we consider.
}, $d{\bar \omega} _i$ -- their complex conjugate, and
$A_{\mu}^i$, ${\bar A}_{\mu}^i$ depend only on the four
$4d$ co-ordinates $x=\{ x^0,x^1,x^2,x^3\} $.

If the metric on $\Sigma _{N_c}$ is chosen so that
$\ast d\omega _i = - d\omega _i$, $\ast d{\bar \omega} _i =
+ d{\bar \omega} _i$, the
self-duality of $C$ implies that the 1-forms $A$
and $\bar A$ in
(\ref{ver}) correspond to the anti-selfdual and selfdual
components of the $4d$ gauge field with the curvature
(tension) $G = \{ G_{\mu\nu}\}$:
\be
dA^i = G^i -{ \ast} G^i
\nn \\
d{\bar A}^i = G^i +{ \ast} G^i
\ee
It remains to subsitute this into (\ref{act}) and use
the relations
\be
\int _{\Sigma _{N_c}}d\omega_i\wedge d{\bar \omega} _j =
2i\Im T_{ij}
\nn \\
\int _{\Sigma _{N_c}}d\omega_i\wedge d{\omega} _j = 0
\ee
where $T_{ij}$ is the period matrix of $\Sigma _{N_c}$
(and depends on the v.e.v.'s of the transverse
scalar fields once the shape of the
curve $\Sigma _{N_c}$ -- its embedding into the
$(x^4,x^5,x^6,x^9)$-space is fixed.
The result for the $4d$ effective action is
\be
\int _{d^4x}\Im T_{ij}G_{\mu\nu}^iG_{\mu\nu}^j +
\hbox{SUSY terms}
\ee
This is essentially the SW answer of refs. \cite{SW1,SW2},
only the part with the topological $\theta$-term is ignored.
It can be restored by more
careful treatment of the self-dual 2-forms:
the action (\ref{act}) is obviously too naive.
There are various approaches developed for this purpose,
see for example refs. \cite{V,W3,Schwarz,PST,West} etc.

Though in principle dictated by supersymmetry,
the analog of expansion (\ref{ver}) is not so trivial
for the scalar fields (the superpartners of the antisymmetric
form in $6d$ and the vector bosons in $4d$).
As usual in the Green-Schwarz formalism on
topologically non-trivial manifolds \cite{KM},
like $\Sigma_{N_c}$, the "embedding matrix"
$\Pi_z = \partial_z\Phi$ is actually
substituted by 1-form on $\Sigma$, with holomorphic
zero modes,
\be
\Pi^{(0)}_z = \sum_{i=1}^{N_c-1} \Pi^i(x) d\omega_i(z)
\ee
This is important for explanation why $N_c-1$ different
scalars emerge from a single $\Phi$, and why the
period matrix of $\Sigma$ appears in the scalar
Lagrangian. Especially transparent should be
(reformulation of) the formalism of ref.\cite{PST},
where appropriate auxiliary field is actually a 1-form
on $\Sigma_{N_c}$, which after gauge fixing becomes
\be
v^{(0)}_z = \sum_{i=1}^{N_c-1} v^i(x) d\omega_i(z)
\ee

\section{Conclusion}

We argued that the recent advance of ref.\cite{W2}
(which reformulated the SW anzatz in the language of branes
and therefore inspired an anzatz for
what the interaction of branes does with the
naive DHW construction)
still does no resolve the basic problem
of all previous considerations: two basic
different ingredients of the SW theory (the spectral curve
and the prepotential) are well justified in two dual pictures.
However, it brings the issue even closer to the main
mysteries of string dualities.

In particular it helps to approach the (still) anticipated
discovery of integrable structures behind the string dynamics.
In this framework one expects extrapolation of the
known results for $2d$, $3d$ and $4d$ models to higher
dimensions.
As to the {\it origin} of the integrability in the theory of
renormalization group flows, it
is a subject of a different investigation.
A possible direction has been suggested
in ref.\cite{GKMMM},
and emerging relation between the
IIB-IIA and the ("nonabelian") Hitchin - ("abelian") Whitham
dualities can provide new insights on this way.

However, even in the restricted framework --
of generalization of the SW theory to
higher dimensions, strings and $M$-theory --
a lot remains obscure.
One of the interesting things to do is to find
the brane analog/interpretation of the mysterious
WDVV-like equations \cite{MMM},
which are peculiar for the
majority of SW effective theories in four and five dimensions,
are related to multiplication "algebra" of 1-forms and
constitute a non-trivial deformation of the WDVV equations for
quantum cohomologies \cite{WDVV}.

\section{Acknowledgements}

We are indebted to many colleagues for valuable discussions
concerning the subject of the present paper, especially to
E.Akhmedov, M.Bianchi, E.Corrigan, A.Gorsky, A.Mironov,
N.Nekrasov, A.Sagnotti, J.Schwarz, M.Tonin and P.West.

A.Marshakov and A.Morozov acknowledge the support of the
Cariplo Foundation and hospitality of the Milano University
and Centro Volta in Como during the work on this paper.

This research was partially supported by the grants
RFBR 96-02-19085 (A.Marshakov)
and RFBR 96-15-96939 (A.Morozov).

\newpage
\vspace{5mm}
\centerline{\hbox{\psfig{figure=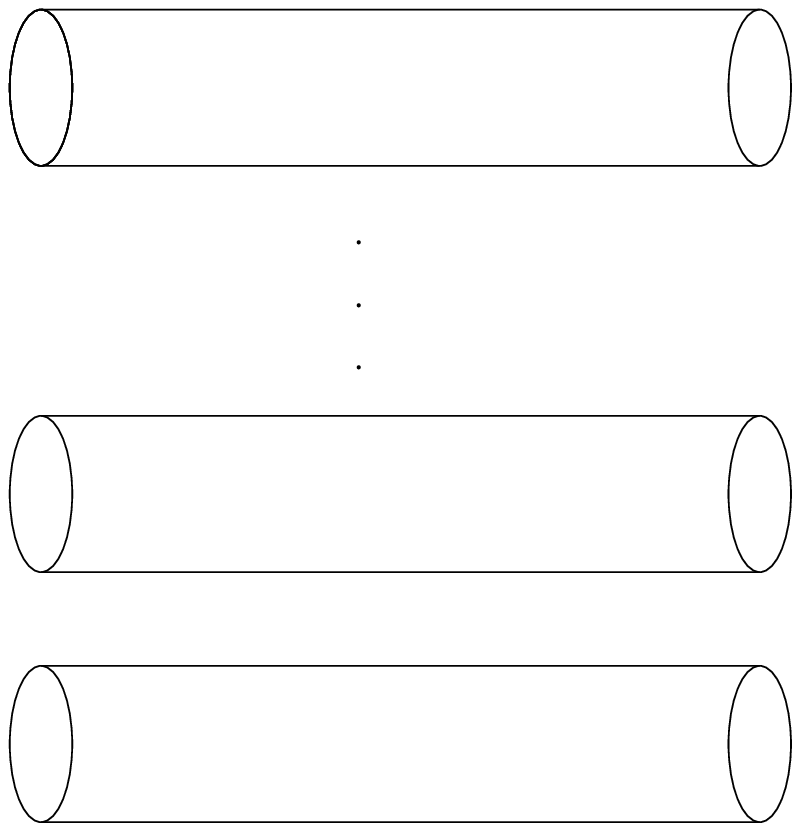}}}
\centerline{Fig.1}
$N_c$ parallel cylinders. The horisontal co-ordinate is
$x^6$, while the vertical axis corresponds to the space
$V^{\bot}$, actually parameterized by $\lambda$.
\vspace{5mm}
\newpage
\vspace{5mm}
\centerline{\hbox{\psfig{figure=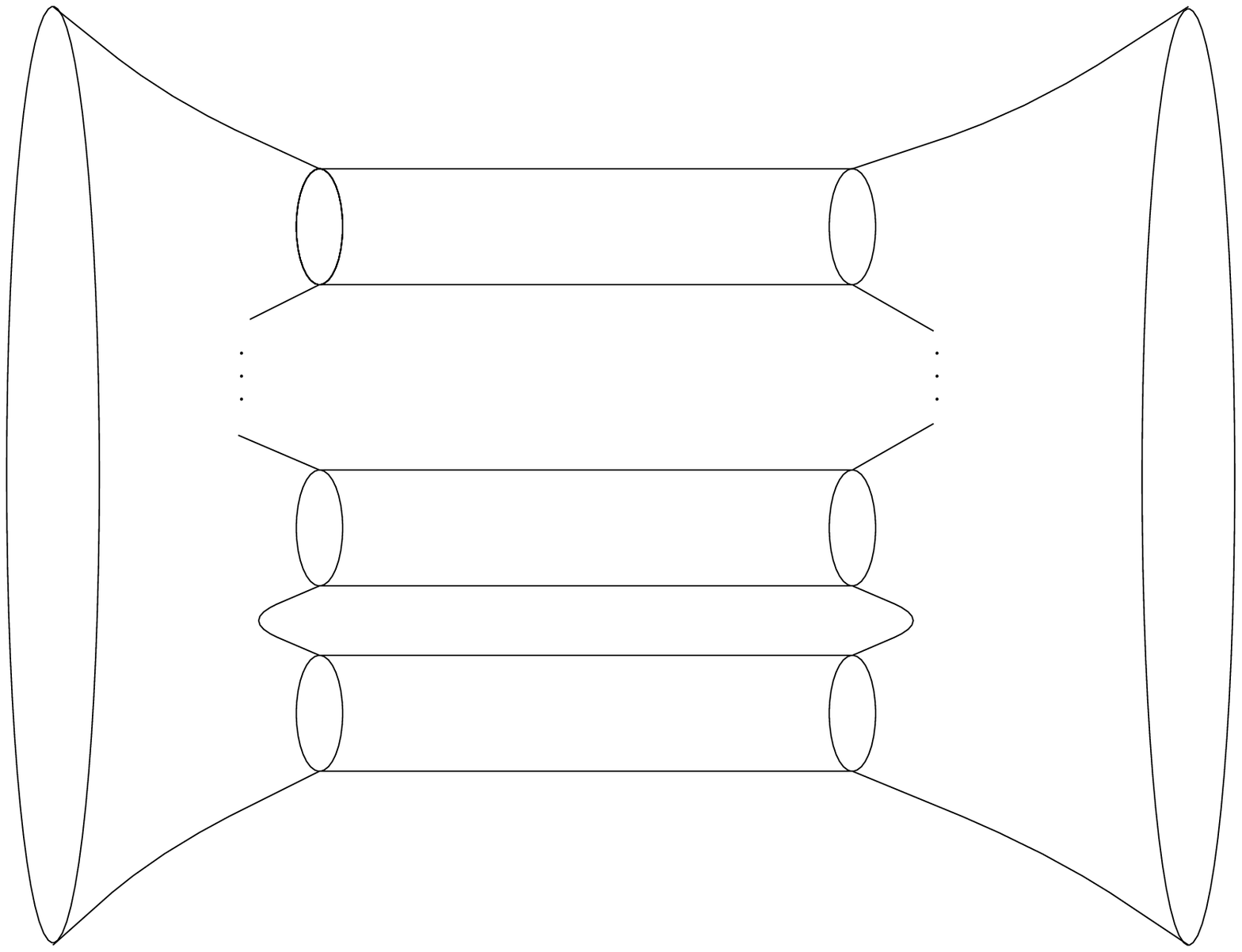}}}
\centerline{Fig.2}
The brane configuration, represented as a result of
gluing $N_c$ cylinders together. Actually, a real-$\lambda$
section of the complex curve (\ref{todacur}) is shown.
The horizontal coordinate is $z$, the vertical one --
$\lambda$. If projected on the vertical plane,
the curve looks like a double-covering of a punctured
Riemann sphere -- the hyperelliptic surface.
If projected on the horizontal cylinder, it is
its $N_c$-fold covering.
\vspace{5mm}
\newpage
\vspace{5mm}
\centerline{\hbox{\psfig{figure=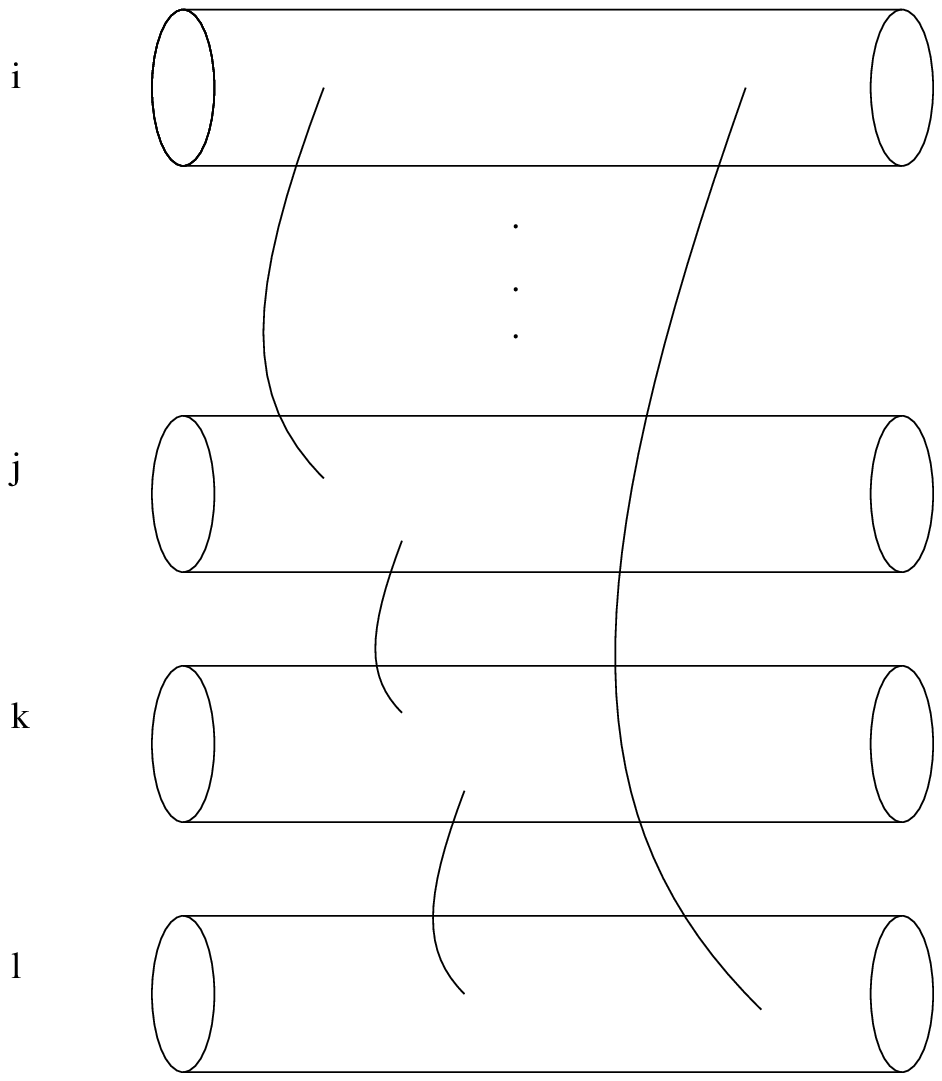}}}
\centerline{Fig.3}
Open strings, stretched between parts of the
$5$-$D$brane. The term in the $6d$ first-quantized
action, associated with this picture is
$\delta ^{M_1...M_4}\int d^6x A^{ij}_{M_1}
A^{jk}_{M_2}A^{kl}_{M_3}A^{li}_{M_4} =
\Tr\left([A_M,A_N]^2\right)$.
\vspace{5mm}
\newpage
\vspace{5mm}
\centerline{\hbox{\psfig{figure=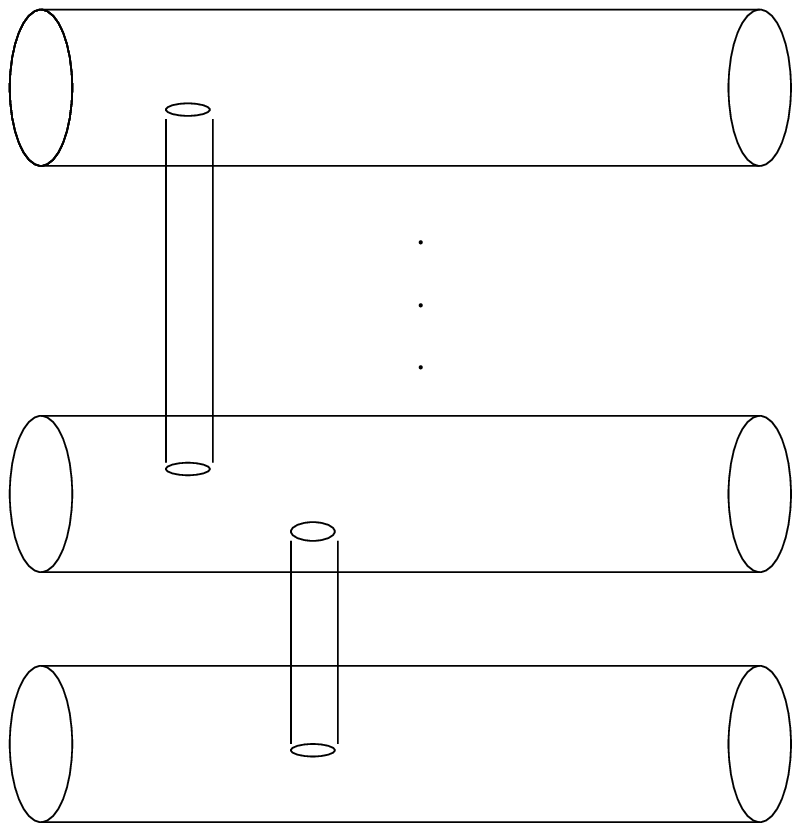}}}
\centerline{Fig.4}
The horisontal cylinders
in Fig.4 are parts of the $5$-brane, as in Fig.1, with 4 flat
dimensions (including the "time"-one $x^0$) not shown on the
picture.
The vertical cylinders on Fig.4 are membranes ($2$-branes) at
a given time, streched between the components of the $5$-brane.
When the width of the vertical cylinders goes to zero, these
membranes turn into the open strings. Moreover, this limit is
not just a result of shrinking some compact dimension -- it is
rather a double-scaling limit,
when the width of the horisontal cylinders remains finite (and
is determined in terms of $\Lambda$ in (\ref{todacur})).
\vspace{5mm}

\end{document}